\documentclass[nofootinbib,prd,twocolumn,showpacs,showkeys,preprintnumbers]{revtex4-1}
\usepackage{hyperref,amssymb,amsmath,mathrsfs,bm,graphicx}
\begin{document}
\title{The mass of a bit of information and the Brillouin's principle}
\author{L. Herrera} 
\email{lherrera@usal.es}
\affiliation{Escuela de F\'\i sica, Facultad de  Ciencias,
Universidad Central de Venezuela, Caracas, Venezuela}

\date{\today}
\begin{abstract}
Based on the Brillouin (negentropic) principle, we   answer to the question: Does information have  mass ?  The obtained answer is affirmative and the mass associated to a bit of information (which is always a positive definite quantity)  is explicitely calculated. Since the Brillouin's principle is modified in the presence of a gravitational field, so does the mass associated to a bit of information. Some consequences ensuing these facts, as well as the link between radiation and information,  are discussed.
\end{abstract}
\date{\today}
\maketitle
\section{INTRODUCTION}
In  recent  years, the question about what could be (if any) the possible mass associated to information,  has arisen in different contexts, and has received many different answers (see \cite{1, 2} and references therein).

We shall not review here the arguments presented in the past to answer to the above mentioned question. Instead we shall provide in this letter an answer based on the Brillouin's principle. Doing that, we are implicitely assuming that the dissipation of energy associated to the change of one bit of information (as implied by  Brillouin's principle) is a fundamental process, independent on the technicalities associated to the information  processing.

 Next we shall analyze the situation when a gravitational field is present, in the context of general relativity.
\section{The Brillouin's principle and the mass of a bit  of information}
According to the Brillouin's principle \cite{3, 4, 5}, the change of  one bit of information stored in a system, requires  the dissipation into the environment (if the temperature of the system remains unchanged) of an amount of energy, whose lower bound is given by 
\begin{equation}
\bigtriangleup E=kT \ln2,
\label{lan1}
\end{equation}
where $k$ is the Boltzmann constant and $T$ denotes the  temperature of the environment. Such an amount of dissipated energy is independent on the details of the  process.

Therefore, according to special relativity, the change of one bit of information (provided the temperature is fixed) leads to  a decreasing of  the mass of the system, by an amount  whose minimal value is:
\begin{equation}
\bigtriangleup M=\frac{kT }{c^2}\ln2,
\label{lan11}
\end{equation}
where $c$ denotes the speed of light.

Since the above quantity determines the decreasing of the mass of the system, associated to the change  of one bit of information, it is fair to say that such a mass is associated to this information.

Thus, for one bit of information, at room temperature, the minimal dissipated energy is 
\begin{equation}
\bigtriangleup E \approx 2.8 \times 10^{-14} erg
\label{lan1bis}
\end{equation}
and  the associated mass is:
\begin{equation}
\bigtriangleup  M\approx 3 \times 10^{-35} grams.
\label{lan111}
\end{equation}

The above figure is about  8 orders of magnitude smaller than the mass of an electron, and about the same order of magnitude as the mass (energy)  associated to an optical photon. 

From (\ref{lan1bis}) a limit in the speed of information processing may be inferred. Indeed, from the uncertainty principle, it follows that the minimal time interval required to measure such an amount of energy is constrained  by
\begin{equation}
\bigtriangleup t\approx\frac{\hbar}{\bigtriangleup E}\approx 3.7\times 10^{-14} s,
\label{la7bis}
\end{equation}
where $\hbar$ is the Planck constant divided by $2\pi$.
In other words, the change  of one bit extends for  $10^{-14} s$. Thus  for all operations involving erasure and/or acquirement of  information, it appears that there exist a limit for the speed processing, which is of the order of $10^5 GHz$.

We shall next consider the case, when the system is placed in a gravitational field.
\section{The mass of a bit  in a gravitational field}

If the system is located in a (weak) static gravitational field, then the Brillouin's principle changes, in a similar way as the Landauer's principle does \cite{lan, pie, Plenio, bais}. Here we have referred to the former, to avoid some  flaws present in the latter (see \cite{6,7, 8} for discussion).

Thus, we have\cite{pla}
\begin{equation}
\bigtriangleup E=kT(1+\frac{\phi}{c^2}) \ln2. 
\label{lan2}
\end{equation}
where $\phi$ denotes the (negative) gravitational potential, and  $T(1+\frac{\phi}{c^2})$ (the Tolman's temperature) is the quantity  which is constant  in thermodynamic equilibrium \cite{Tol}.

The very idea underlying the concept of Tolman's temperature is very simple indeed, namely: since according to special relativity all forms of energy have inertia, this should also apply to heat. Therefore, because of the equivalence principle, there should be also some weight associated to heat, and one should expect that thermal energy tends to displace to regions of lower gravitational potential. This in turn implies that the condition of thermal equilibrium in the presence of a gravitational field must change with respect to its form in  absence of gravity. Thus a temperature gradient is necessary in thermal equilibrium in order to prevent the flow of heat from regions of higher to lower gravitational potential. This result was confirmed some years later by Eckart and Landau and Lifshitsz \cite{Eckart, L}. Indeed, in the transport equation derived by these authors, the ``inertial'' term deduced by Tolman appears explicitly.

In the case of  a field of arbitrary strength (in general relativity),  Tolman's temperature becomes $T\sqrt{g_{tt}}$, accordingly   (\ref{lan2})  generalizes to \cite{LH}
\begin{equation}
\bigtriangleup E=kT \sqrt{g_{tt}}\ln2,
\label{lan3}
\end{equation}
producing
\begin{equation}
\bigtriangleup M =\frac{kT}{c^2} \sqrt{g_{tt}}\ln2,
\label{lan4}
\end{equation}
where $g_{tt}$ denotes the $tt$ component of the metric tensor (the coefficient of $dt^2$ in the expression for the line element).

From (\ref{lan3}), it follows    that at the horizon (the infinite redshift surface, where  $g_{tt}=0$ and the source of the gravitational field  becomes a black hole), either the proper temperature becomes singular or the change  of information can be done without any dissipation of energy. If we exclude the  former possibility on physical  grounds, then we are left with the fact that $\bigtriangleup E$ should vanish at the horizon, leading to a vanishing mass for a bit of information.

Now, as it follows from the information theory \cite{pie}, this situation (change of information  without dissipation) corresponds to the case  where all  bits are already in one state only. This is exactly the situation that emerges from the assumption that the quantum radiation  emitted by the black hole   is nearly thermal \cite{H1, H2}, which in turn suggests the ``bleaching'' of information at the horizon. 

To summarize: once the system is surrounded by a horizon ($g_{tt}=0$), no further information  leaves the system (a known fact, but now deduced from information theory). Otherwise, if $g_{tt}\neq 0$, the corresponding mass of a bit of information decreases. For the case of the earth, though, the corresponding decrease is extremely small ($\frac{\phi}{c^2}\approx 10^{-9}$).
\section{Information and radiation}
Finally we would like to call the attention to a question, tightly related to the issue discussed here, about the link between the energy and the information, conveyed by radiation. 

In fact, by means of radiation (at classical level), the source of the field ``informs'' about any changes in its structure (this includes of course changes in its state of motion).

Thus, the knowledge of the field variables (as well as their derivatives of any order) outside the source, on a given hypersurface (spacelike or null), is not sufficient to forecast, {\it modulo} the field equations, the future of the field beyond that hypersurface. This is particularly well illustrated in the Bondi formalism \cite{bondi, 17, 18} .

The information required to forecast the evolution of the system (besides the ``initial'' data) is identified  with radiation itself. In the context of the  Bondi formalism this information is represented by the so called ``news function''. It appears that, whatever happens at the source, leading to changes in the field, it can only
do so by affecting the news function and vice versa. Thus the relationship between the news function and the occurrence of radiation becomes clear. This scheme applies to Maxwell systems in Minkowski spacetime \cite{janis} as well as to Einstein--Maxwell systems \cite{18}.

Furthermore, it turns out that the Bondi mass of a system (a measure of its total mass),  is constant if and only if there are no news. 

From the above comments it should be clear that the emitted radiation conveys information and mass (energy), a well known fact. On the other hand a bit of radiated information  implies a bit of erased information at the radiating system, leading to  a decrease of its total mass (energy). This fact illustrates further the idea that Brillouins principle provides a bridge between information theory and physics.
 If we put  together all these pieces, we are lead to the following  question: what part of the total radiated energy (mass)  corresponds to the radiated information? Obviously the answer depends on the type of radiation under consideration, being zero for  the case of thermal radiation. Notwithstanding, we believe that this issue, revealing the relationship between information and radiation is a  very relevant one, and  deserves further attention.
\section{Summary}
This letter addresses the problem about the existence of mass associated with information. The novelty in our approach consists in that we base  the answer on a fundamental principle of information theory (Brillouin's principle). Doing so, our result is independent of any kind of technicalities regarding the acquiring, erasing and processing of the information.
 We provide an explicit value for one bit of information. Since the Landauer's principle is modified when the system is placed in a gravitational field, we extend our result to this latter case too. Finally we briefly discuss about 
 the link between radiation and information.
\begin{acknowledgments}
I would like to thank Prof. Kish, for many useful comments regarding the Brillouin's principle.
\end{acknowledgments}

\end{document}